\documentclass[10pt, a4paper]{article}

\usepackage[utf8]{inputenc}
\usepackage{graphicx}
\usepackage[margin=1in]{geometry}
\usepackage{subcaption}
\usepackage{authblk}
\usepackage{amsmath}
\usepackage{verbatim}
\usepackage{lipsum}
\usepackage{mathtools}
\usepackage{cuted}
\usepackage{xcolor}
\usepackage{amssymb}

\begin{document}

\title{A novel analytical formulation of the Axelrod model}

\author[1,2]{Luc\'ia Pedraza}
\author[1,2]{Sebasti\'an Pinto}
\author[3]{Juan Pablo Pinasco}
\author[1,2]{Pablo Balenzuela}

\affil[1]{\small{Departamento de F\'isica, Facultad de Ciencias Exactas y Naturales, Universidad de Buenos Aires. Av.Cantilo s/n, Pabell\'on 1, Ciudad Universitaria, 1428, Buenos Aires, Argentina.}}
\affil[2]{\small{Instituto de F\'isica de Buenos Aires (IFIBA), CONICET. Av.Cantilo s/n, Pabell\'on 1, Ciudad Universitaria, 1428, Buenos Aires, Argentina.}}
\affil[3]{\small{Departamento de Matemática, Facultad de Ciencias Exactas y Naturales, Universidad de Buenos Aires and IMAS UBA-CONICET, Av. Cantilo s/n, Pabellón 1, Ciudad Universitaria, 1428, Buenos Aires, Argentina.}}

\maketitle

\begin{abstract}
\par The Axelrod model of cultural dissemination has been widely studied in the field of statistical mechanics. The traditional version of this agent-based model is to assign a cultural vector of $F$ components to each agent, where each component can take one of $Q$ cultural trait. In this work, we introduce a novel set of mean field master equations to describe the model for $F=2$ and $F=3$ in complete graphs where all indirect interactions are explicitly calculated. We find that the transition between different macroscopic states is driven by initial conditions (set by parameter $Q$) and the size of the system $N$, who measures the balance between linear and cubic terms in master equations. We also find that this analytical approach fully agrees with simulations where the system does not break up during the dynamics and a scaling relation related to missing links reestablishes the agreement when this happens.

\end{abstract}

\section{Introduction}

\par The Axelrod model \cite{axelrod1997dissemination} has been proposed to explain the phenomenon of polarization in a a society in which individuals always are looking for a consensus in their opinions. This model is based on two well-established mechanisms: Social influence, through which people become more similar when they interact; and homophily, which is the tendency of individuals to interact preferentially with similar ones.
The mathematical description of this model consists on describing individuals as agents, each one described by a vector of $F$ components called cultural features which at the same time can take one of $Q$ integer values called cultural traits. When two agents interact with a probability proportional to their shared features, one of them copies a feature from the other one.
Despite the simplicity of the model and beyond its original formulation, several variants are still being proposed in order to improve the modeling of polarization, such as the emergence of new topics \cite{hernandez2018robustness}, a layered organization of social interactions \cite{battiston2017layered}, the inclusion of peer-pressure \cite{morris2019impact}, and the formation of opinion-based groups \cite{maccarron2020agreement}.

\par Which is interesting from a statistical physics point of view is that this model shows a non-equilibrium phase transition from a monocultural to a multicultural state. This phase transition takes place by varying the number of cultural traits $Q$ for a given fixed $F$. If the number of cultural traits is low, the probability of interaction is high, leading the system to a monocultural state.  On the other hand, if $Q$ is high, the mentioned probability is low and the system evolves to a stationary multicultural state after a few interactions.
This phase transition was studied in several topologies such as one-dimensional systems \cite{klemm2005globalization, lanchier2013fixation}, lattices \cite{castellano2000nonequilibrium},  complex  \cite{klemm2003nonequilibrium} and complete networks \cite{pinto2020erdos}.

\par Although the Axelrod model is usually studied through numerical simulations, different analytical approaches were developed based on stochastic equations \cite{castellano2000nonequilibrium, stivala2016another} or deterministic systems \cite{vilone2002ordering, vazquez2007non}, that describe the evolution of the density of bonds with a given similarity and provide some insights about the origin of the phase-transition. However, these approaches rely on several approximations needed to discard high-order terms and make equations analytical tractable. Moreover, the more fully descriptive works are necessarily only devoted to the simplest case $F=2$ \cite{vazquez2007non, lanchier2016fixation}.

\par In this work, we develop a new set of mean field equations on complete networks that exactly describe the average behavior of the Axelrod model in the $F = 2$ case and provide a full description of the similarity distribution dynamics for larger $F$. These equations are based on a novel formulation of the model in terms of similarity vectors among agents. This formulation naturally takes into account correlations among cultural states and simplifies the model due to the parameter $Q$ is only involved in setting the initial condition.
We show that the  $F=2$ case reduces to a trivial dynamical behavior, while the case $F=3$ shows a competition between linear and cubic terms mediated by the size of the system $N$. In this last case, the analytical approach shows a fully agreement with simulations when $N$ increases at fixed $Q$ (below the transition point), but fails during the transition where the mean-field hypothesis do not hold. However, the agreement is rapidly recovered by a scaling factor related to missing links due to the system fragmentation during the dynamics.

\section{Axelrod model}

\par The Axelrod model \cite{axelrod1997dissemination} describes each agent by a vector of $F$ components which can take one of $Q$ integer values.
That vector represents a set of cultural features associated to a given individual, and the different values a component adopts represent different cultural traits related to a given feature.
The model starts by creating random cultural states for the agents. In this work, the initial state is set by assigning with equal probability one of $Q$ integer values to each cultural feature.
Once the initial condition is established, the dynamics of the system is based on a pairwise interaction mechanism, which relies on two fundamental hypothesis:
\begin{itemize}
\item Homophily: The probability of interaction between two topological connected individuals is proportional to their cultural similarity, that is, the number of features they share. More specifically, two agents interact with probability $n/F$, where $n$ is the number of shared features. If $n = F$, the agents do not interact.
\item Social Influence: After each interaction, the agents become more similar. It means that one of the agents copies a feature from the other which they previously did not share.
\end{itemize} 
The system evolves until there are no active links in the system, i.e., all connected agents either do not share any feature or share $F$ cultural features.
\par This model shows a non-equilibrium phase transition from a monocultural to a multicultural state by varying the value of $Q$ for a fixed $F$. When $Q < Q_c$, the probability that two agents can interact since the initial state is high, so all agents end with the same cultural vector.
On the other hand, when $Q > Q_c$, the probability of interaction at the initial state is low and the final state shows a coexistence of regions with different cultural states.
The transition point and the order of the phase-transition depends on the values of $F$ and on the topology of the underlying contact network. In this work, we study the model on a complete network, where a given agent can interact with any other agent in the system.

\subsection{Vector description of similarities}
\label{sec:vector_similarity}.

\par
\begin{figure}[ht]
    \centering
    \includegraphics[width = \columnwidth]{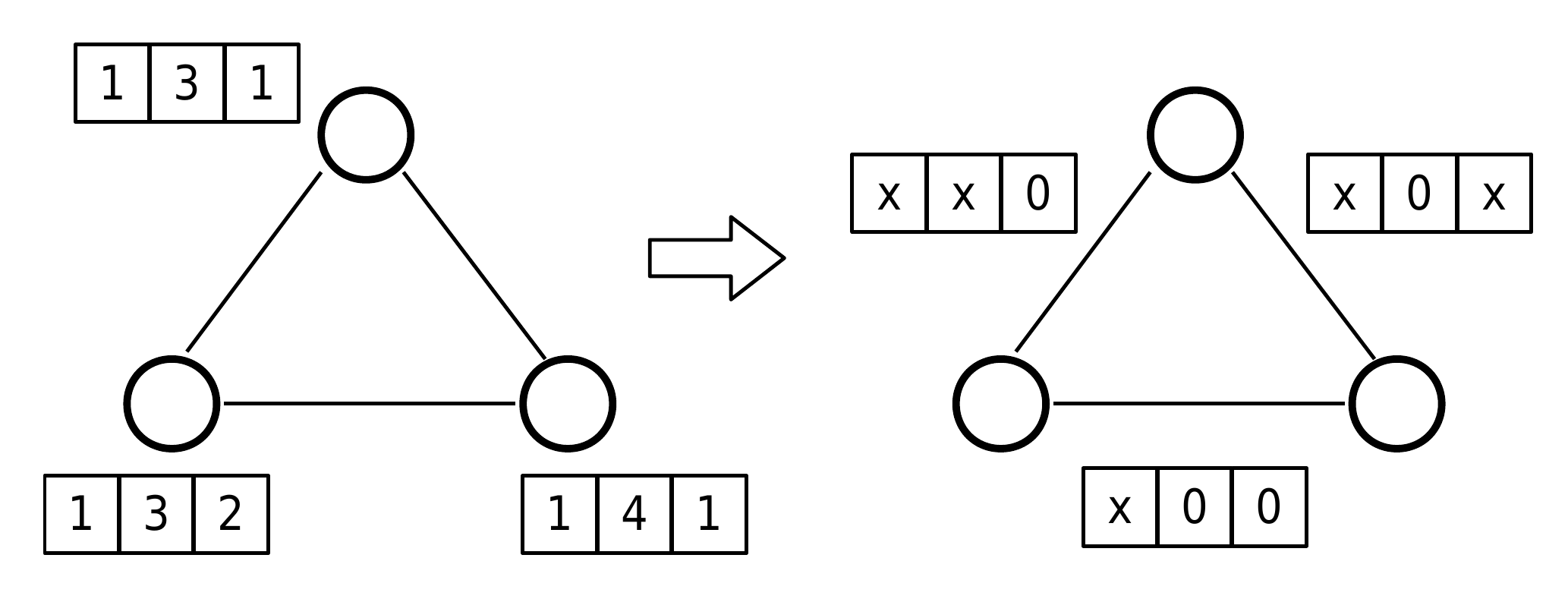}
    \caption{{\bf Axelrod model described in terms of similarity vectors.} Cultural vectors (left figure) of $F$ components and $Q$ integer values per feature associated with each agent are replaced by binary similarity vectors (right figure) of $F$ components associated to each pair of agents.}
    \label{fig:Similarity_vectors}
\end{figure}

 Here we postulate an alternative formulation of Axelrod model based on link's dynamics instead of agent's dynamics. Given than similarity between agents plays a main role in the dynamics of the model, we  introduce this new formulation  in terms of similarity vectors. This framework will allow us to write closed mean field master equations for describing the dynamics of the model, as we shown in next sections. 
Figure \ref{fig:Similarity_vectors} shows how the description based on individual cultural vectors is seen in terms of similarity vectors associated to any pairs of agents. This formulation implies to describe the system of $N$ agents, with their original $N$ cultural vectors, in terms of $\frac{N(N-1)}{2}$ similarity vectors. 
\par A similarity vector between two agents is a F-dimensional binary vector which has an $X$ in the place where they share a cultural feature and $0$ otherwise. Both formulation are identical (see Appendix for details) if they fulfill the following relationships: 
Given a state of the similarity vector between agent $i$ and $j$ ($i-j$), if some feature adopts the value of $X$, then the respective feature in the similarity vectors $i-k$ and $j-k$ must be equal ($0$ or $X$) for all $k$ closing the triangle $i-j-k$. Any change in a similarity vector must fulfill this condition. (A full description of the dynamics is sketched in the Appendix). 
 
\par The formulation in terms of similarity vectors shows that the parameter $Q$ only plays a role in setting the initial state: A similarity vector of length $F$ and $n$ $X$-values appears at the initial state with a probability:
\begin{equation}
    P(n | F, Q) = \binom{F}{n}(\frac{1}{Q})^n (1 - \frac{1}{Q})^{F-n}
    \label{eq:initial_conditions}
\end{equation}
Once the initial state is set, the absolute value of $Q$ is meaningless from the point of view of the dynamics. This independence of $Q$ is already present in the mean-field approach given in \cite{castellano2000nonequilibrium}, but it seems to be lost in most of the Axelrod literature.

\section{Analytical formulation}

\par The formulation in terms of similarity vectors allows to derive mean-field equations for the similarity distribution  in terms of the density of states. 
We introduce here the analytical approach for the cases $F = 2$ and $F = 3$. 

\subsection{Case $F = 2$}

\par In this case, a given link can be in four states: $P_0=[0,0]$, $P_{1a}=[X,0]$, $P_{1b}=[0,X]$ and $P_2=[X,X]$. These states can not coexist in a given trio of agents due to the closure relationship detailed in previous section.  Moreover, given the homophily driven pairwise interaction, a change in the cultural state of one agent due to a direct interaction with his partner produces not only an update in the similarity of the current link but also indirect updates in the similarities of all the other pairs of agents which involve the former one. 

\par
\begin{figure}[ht]
    \centering
    \includegraphics[width = \columnwidth]{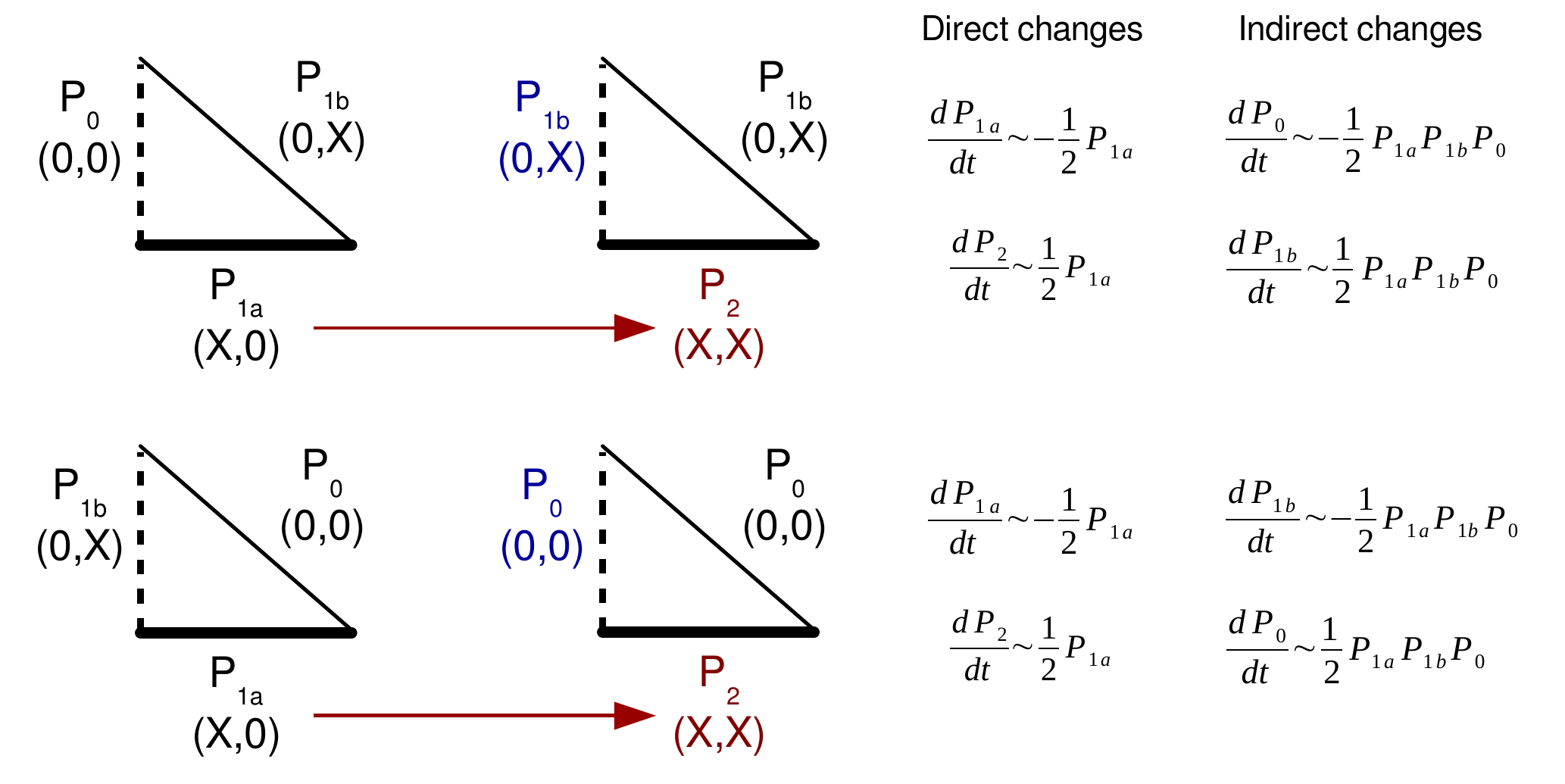}
    \caption{{\bf Example of dynamics of vector similarity states.}
    Two examples of links updates which involves direct and indirect interactions. Note that direct changes lead to linear terms in Eq.(\ref{eq:F2}), and these direct changes imply indirect ones, pointed out by dashed lines, which at the same time lead to cubic terms
    The $1/2$ factor corresponds to the probability that a direct change effectively occurs (i.e., $1/F$ for $F=2$ and one cultural feature shared).}
    \label{Esquema_F2}
\end{figure}

 Figure \ref{Esquema_F2} shows two examples of direct and indirect link updates that give rise to some terms of the equations. Bold lines represent those links where a direct change can take place, and correspond to states $P_{1a}$ or $P_{1b}$ because direct changes do not take place if similarity is zero ($P_0$) or one ($P_2$). If we consider all possible combinations and direct changes, these lead to the following equations for the similarity vector states, where direct changes produce linear terms and indirect changes are reflected in cubic terms in the equations:

\begin{align*}
    \frac{dP_0}{dt}&=\frac{(N-2)}{4}\Big(- P_{1a}P_{1b}P_0 - P_{1a}P_{1b}P_0 + P_{1a}P_{1b}P_0 + P_{1a}P_{1b}P_0\Big) \\
    \frac{dP_{1a}}{dt}&=-\frac{P_{1a}}{2}+\frac{(N-2)}{4}\Big(-P_{1a}^2P_2-P_0P_{1a}P_{1b}+P_0P_{1a}P_{1b}+P_{1a}^2P_2\Big)\\
    \frac{dP_{1b}}{dt}&=-\frac{P_{1b}}{2}+\frac{(N-2)}{4}\Big(-P_{1b}^2P_2-P_0P_{1a}P_{1b}+P_0P_{1a}P_{1b}+P_{1b}^2P_2\Big)\\
    \frac{dP_2}{dt}&=\frac{P_{1b}+P_{1a}}{2}+\frac{(N-2)}{4}\Big(-P_{1a}^2P_2-P_{1b}^2P_2+P_{1a}^2P_2+P_{1b}^2P_2\Big),
\end{align*}
where the $(N-2)$ factor represents the amount of indirect links that might change, considering a complete graph. As we can see in these last equations, the indirect changes effect is canceled and only the direct dynamic remains. 
By calling $P_1$ to the density of links in states $P_{1a}$ or $P_{1b}$, and considering that both $a$ and $b$ type are equally probably, we finally obtain a set of equations for $F = 2$ that can be explicitly solved:
\begin{equation}
\begin{split}
    \frac{dP_0}{dt} = 0 &\implies P_0(t) = P_0(t = 0)\\
    \frac{dP_1}{dt} = -\frac{P_1}{2} &\implies P_1(t) = P_1(t = 0) e^{-\frac{1}{2}t}\\
    \frac{dP_2}{dt} = \frac{P_1}{2} &\implies P_2(t) = 1 - P_0(t = 0) - P_1(t = 0)e^{-\frac{1}{2}t}
\end{split}
\label{eq:F2}
\end{equation}
These equations fulfill the normalization constraint $P_0 + P_1 + P_2 = 1$, and the initial condition is given by Eq.(\ref{eq:initial_conditions}).
\par
\begin{figure}[ht]
    \centering
    \includegraphics[width = \columnwidth]{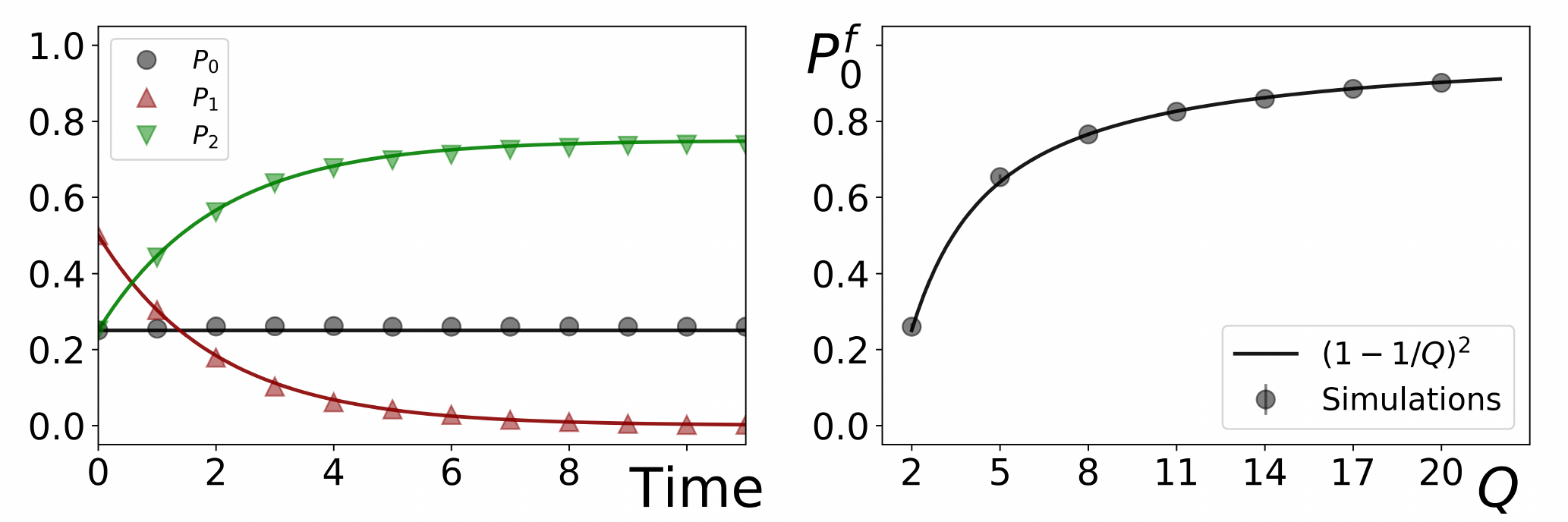}
    \caption{{\bf Analytical prediction for $F = 2$.} 
    Time evolution of similarity distribution for $F = 2$ and $Q = 2$ (panel (a)), and $P_0$ at the final state (panel (b)). Dots belong to simulations while lines are analytical predictions. $N = 256$, but no dependence on $N$ was observed. Time is measured in $\frac{1}{M}$ interactions, where $M =\frac{N(N-1)}{2}$ the number of pair of agents in the system.}
    \label{fig:F2}
\end{figure}

 Figure \ref{fig:F2} shows the comparison between equations (\ref{eq:F2}) and the average of the similarity distribution in a complete graph Axelrod model.
In this figure, we can see the fully agreement between analytical and simulation results: Equations (\ref{eq:F2}) correctly predict that $P_0$ remains constant during the whole dynamics and equal to its initial value $(1-1/Q)^2$, while $P_1$ and $P_2$ exponentially decay.
On the other hand, we do not observe a dependence on $N$ once the time scale is adjusted as $dt = \frac{1}{M}$ where $M = \frac{N(N-1)}{2}$.

\par The conservation of $P_0$ provides a picture of the Axelrod dynamics in an average sense, which can be visualized as a rearrangement of non-zero similarity links in connected components that are also cliques in the final state. 
Finally, we want to remark that the dependency on $Q$ (as stated above) is only present in the initial similarity distribution ($P_i(t=0)$) and these parameter plays no role during the dynamics.

\subsection{Case $F = 3$}

\par Following the same approach detailed above, we write down the dynamical equations for the similarity vectors when $F = 3$ (See Appendix for details).
In this case, we obtain:
\begin{equation}
\begin{split}
    \frac{dP_0}{dt}&=\frac{(N-2)}{27}\Big(P_1^3-3 P_0 P_1 P_2\Big)\\
    \frac{dP_1}{dt}&=-\frac{P_1}{3}+\frac{(N-2)}{27}\Big(-2  P_1^3 -3 P_1^2 P_3 +P_1 P_2^2  +6 P_0 P_1 P_2  \Big)\\
    \frac{dP_2}{dt}&=\frac{P_1}{3}-\frac{2 P_2}{3}+\frac{(N-2)}{27}\Big(P_1^3+6 P_1^2 P_3-2 P_1 P_2^2 -3P_0 P_1 P_2\Big)\\
    \frac{dP_3}{dt}&=\frac{2 P_2}{3}+\frac{(N-2)}{27}\Big(-3 P_1^2 P_3 + P_1 P_2^2 \Big)
\end{split}
\label{eq:F3}
\end{equation}
Due to the normalization condition $P_0+P_1+P_2+P_3=1$, this equation system is actually a three-dimensional one. As in the previous case, the initial condition is given by Eq.(\ref{eq:initial_conditions}).

\par The equation system (\ref{eq:F3}) has a set of fixed points at $P_1 = P_2 = 0$ which corresponds to the case when there are no active links in the system.
This condition does not impose any constraint on $P_0$ and $P_3$. Therefore, the stationary state is fully characterized by specifying the value of $P_0$ at the final state, $P_0^f$, which in principle can be any value between $0$ and $1$. On the other hand, $P_3^f$ is determined through the normalization condition $P_3^f = 1 - P_0^f$. The eigenvalues of the linearized equations are $\{0,-1/3,-2/3,0\}$, which confirm that the infinite fixed points are stables (see Appendix for details). The equation system (\ref{eq:F3}) has also an isolated fixed point which explicitly depends on $N$, but it is an unfeasible solution due to some $P_i$ fall out of the range $[0,1]$ and therefore is discarded for future analysis. 

\par
\begin{figure}[ht]
    \centering
    \includegraphics[width = \columnwidth]{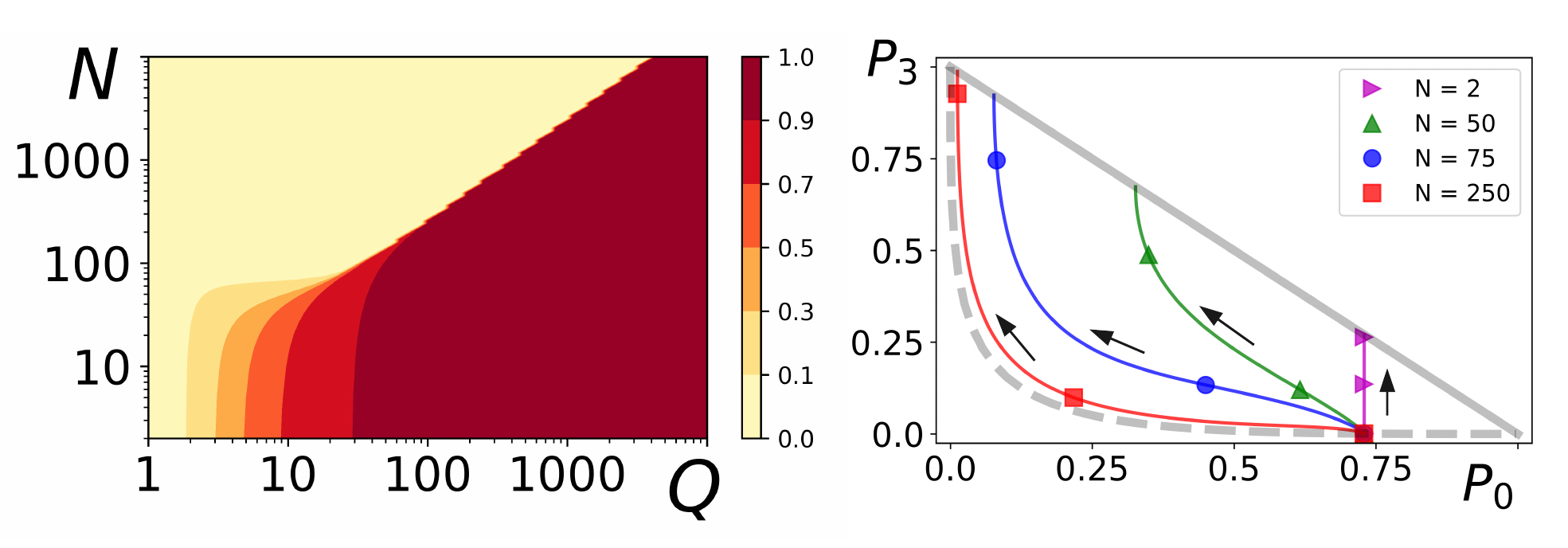}
    \caption{{\bf Phase diagrams of the analytical system.} Stationary solution $P_0^f$ as a function of $N$ and $Q$ (left panel), and phase diagram for different $N$ with same initial condition ($Q=10$) (right panel). In the last one, the dashed line is the set of initial conditions, the solid line is the set of fixed points, and arrows point out time direction.}
    \label{fig:phase_diagram}
\end{figure}

 To which of the final values of $P_0$ the system converges depends on the parameters $Q$ and $N$, as the left panel of figure \ref{fig:phase_diagram} shows. Here we can observe that for small values of $N$ (roughly $N<100$), the system moves continuously from $P_0^f=0$ to $P_0^f=1$ as $Q$ increases, while $P_0^f$ shows an abrupt jump from $0$ to $1$ when $N$ is roughly greater than $100$. This jump takes place at a critical value of Q which scales linearly with $N$, following the relationship $Q_c/N \simeq 0.4$. 

\par Although $P_0^f$ depends on $Q$ and $N$, these parameters play two well-separated roles.  In one hand, $Q$ determines the initial conditions of the system according to Eq. (\ref{eq:initial_conditions}), while on the other hand, $N$ weights the coupling of the cubic terms with the linear ones.
Before going on the analysis of the entire system, it is interesting to look at the case of large $N$: Here, we could neglect the linear terms respect to the cubic ones, leading to a new  equation system without linear terms. This approximation presents a set of stable fixed points with a predominant attractive component which fulfills the following relationships: $P_1=(1-\sqrt[3]{P_0})(\sqrt[3]{P_0})^2$  $P_2=(1-\sqrt[3]{P_0})^2\sqrt[3]{P_0}$ and $P_3=(1-\sqrt[3]{P_0})^3$ (the stability analysis of this approximation can be found in the Appendix).

\par Given that $P_1^f = P_2^f = 0$, in panel (b) of figure \ref{fig:phase_diagram} we analyze the trajectories of the system in the $P_0$-$P_3$ plane for different values of $N$ at fixed $Q$.
Here, the straight line of slope $-1$ is the set of stable points of the whole system, while the dashed gray line is the set of stable fixed points of the cubic part (the  cubic-stable manifold). What is interesting here is that the relation that defines the cubic-stable manifold, $P_3=(1-\sqrt[3]{P_0})^3$, is satisfied by $P_0$ and $P_3$ at the initial condition given by the Axelrod model (Eq.( \ref{eq:initial_conditions})). This means that, at $t = 0$, the cubic terms of equations (\ref{eq:F3}) become equal to zero and the dynamics is driven only by the linear terms (See Appendix for details).

\par In panel (b), the four trajectories sketched ($Q=10$ and $N=2; 50; 75; 250$) display how the balance between linear and cubic terms rules the dynamics: When $N$ is small, the linear term is dominant and the system evolves to the closest fixed point in the straight line, as is shown in the case $N=2$ where cubic terms are absent. However, when $N$ is large (for instance, $N=250$), the linear terms drives the dynamics at $t=0$, but once the system leaves the dashed gray line, these are negligible with respect to the cubic ones, and the system is driven by the stability of that curve. Then, the trajectory goes to the stability points in the straight line, but following the dashed gray line as can be seen in the figure (red trajectory in panel (b)). In this example, for intermediates values of $N$, the dynamics is driven by the competition between the linear and cubic terms. For larger values of $Q$, the system starts from regions of higher values of $P_0$ and either jumps to the closest stable states on the straight line (which also means high values of $P_0^f$) for small values of $N$, or moves following the cubic-stable manifold until reaching the stable state at low values of $P_0^f$ for large values of $N$.

\par
\begin{figure}[ht]
    \centering
    \includegraphics[width = \columnwidth]{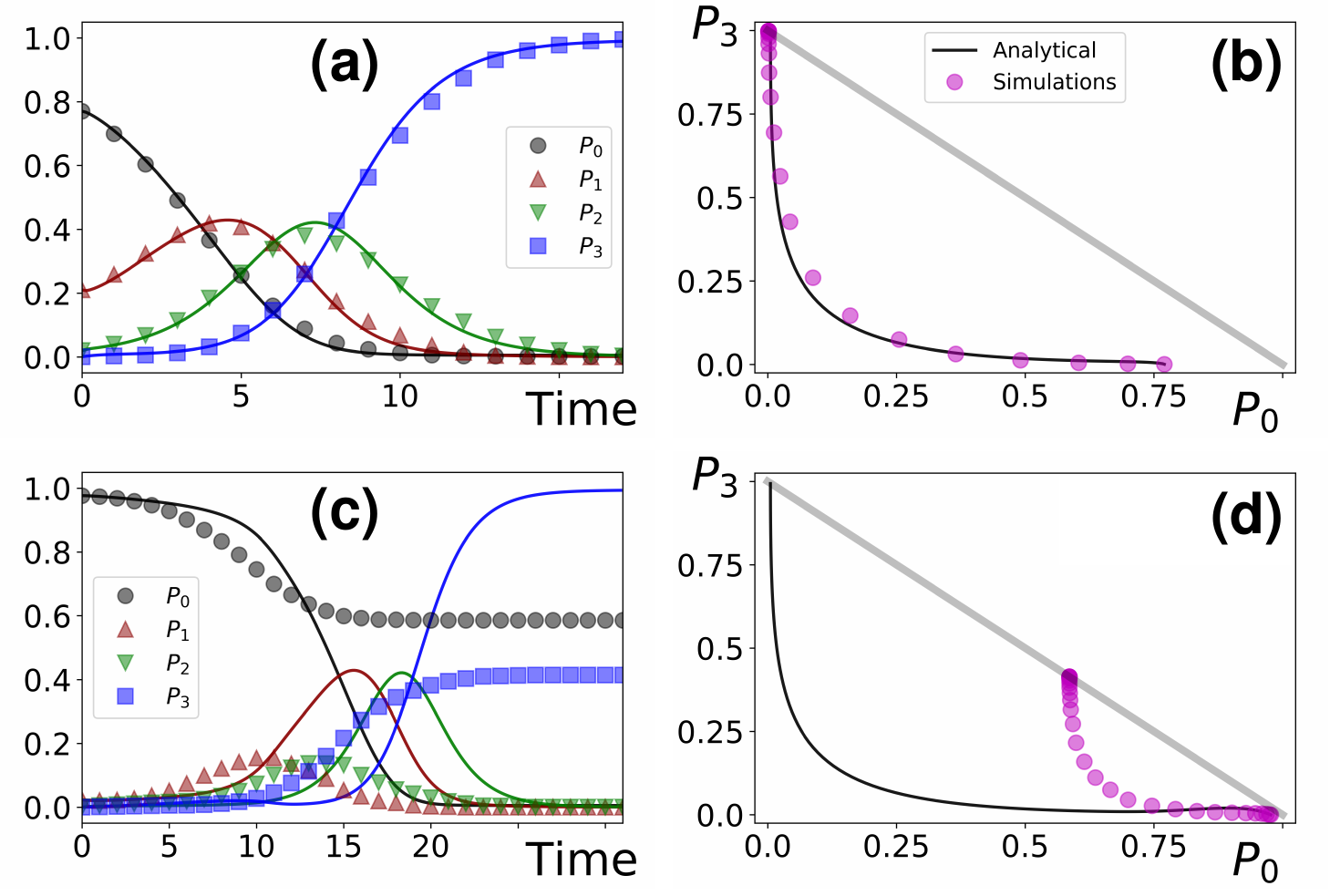}
    \caption{{\bf Phase diagrams or Time evolution for $F = 3$, $N = 512$ and $Q = 12$ (panels (a) and (b)) and $128$ (panels (c) and (d)).} Dots belong to simulations while lines belong to the analytical approach. Panels (a) and (c) show the time evolution of the similarity distribution, while panels (b) and (d) show the same information as phase diagrams.}
    \label{fig:F3_dynamics}
\end{figure}

 The comparison between the analytical approach and simulations can be observed in Figure \ref{fig:F3_dynamics}. In contrast to $F = 2$, the $F = 3$ case shows a dependence on $N$ and the matching between analytical equations and simulations improves when $N$ increases for finite $Q$.
When this happen, $P_0$ decays to zero leading to a monocultural state in the thermodynamic limit for finite $Q$ ($Q/N \to 0$). However, Figure \ref{fig:F3_dynamics} also shows an example where the analytical solution does not follow the Axelrod dynamics and, as we will see, this happens during the Axelrod transition. The comparison between this transition and the observed in the analytical system at $Q/N \sim 0.4$ is provided in the following section.

\subsubsection{Axelrod phase transition in the analytical model.}

\par
\begin{figure}[ht]
    \centering
    \includegraphics[width = 0.80\columnwidth]{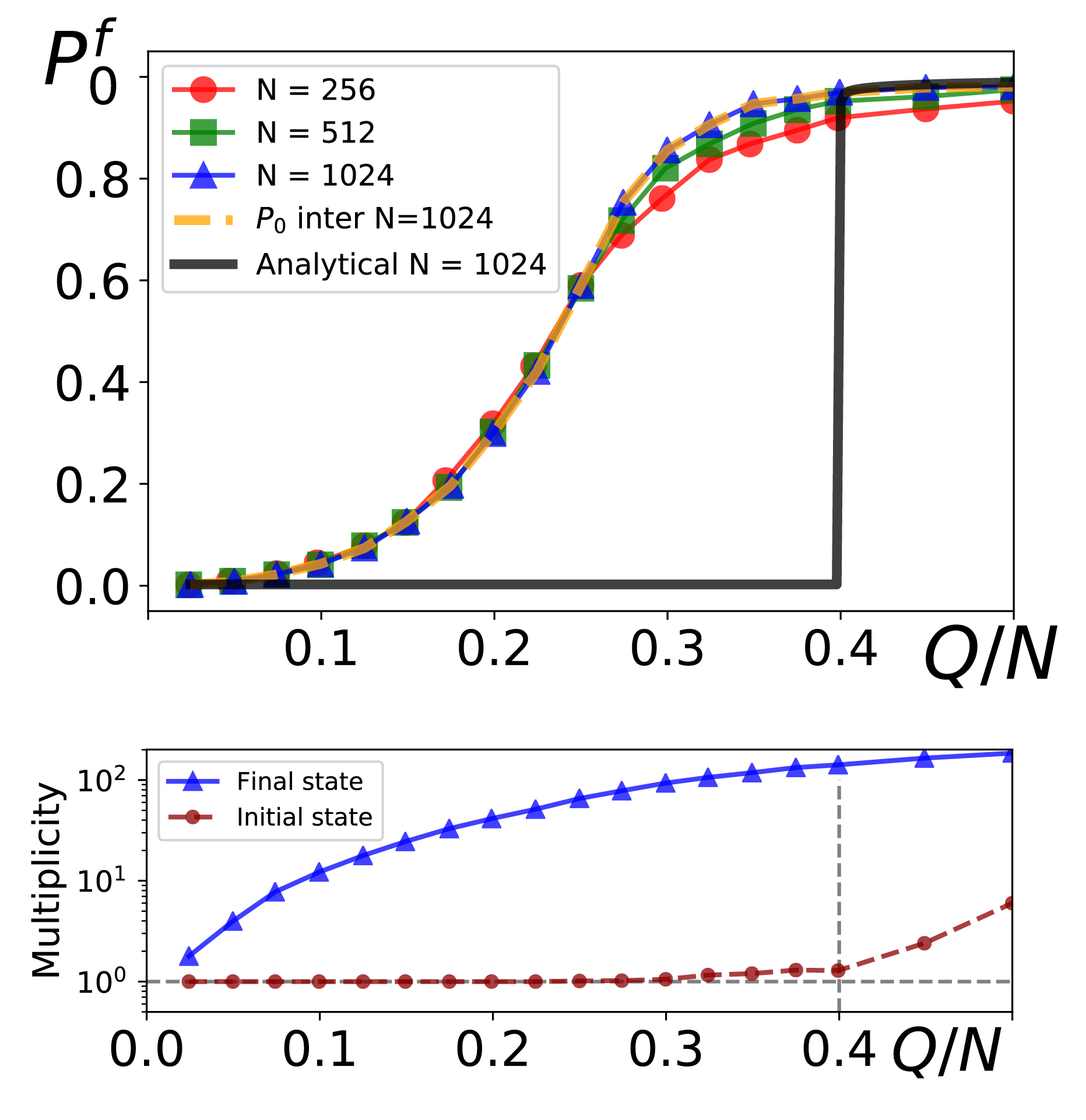}
    \caption{{\bf Axelrod and analytical transition in terms of $P_0^f$.} Top right panel shows the transition for different $N$.
    Bottom right panel show the multiplicity (number) of fragments at the initial and final state of the Axelrod model for $N = 1024$. It can be seen that the analytical transition corresponds to an initial fragmented system, while the difference between the Axelrod model and the mean-field approach corresponds to a fragmented system at the final state.
    }
    \label{fig:transicion}
\end{figure}

 In terms of the similarity distribution, the phase transition of the Axelrod model (the passage from a monocultural state to a multicultural state when $Q$ increases) corresponds to a change in $P_0$ from zero to a non-zero value at the final state ($P_0^f$). Top panel of figure (\ref{fig:transicion}) shows this transition together with the predicted value of the analytical approach as a function of $Q/N$. This scaling was in part suggested by the phase diagram in panel (a) of figure (\ref{fig:phase_diagram}) . 
As top panel of Figure \ref{fig:transicion} shows, the analytical approach differs respect to the Axelrod model during the transition, but it matches the simulations for low values and high values of $Q/N$. In this figure, it is clearer that the analytical system shows a critical value equal to $Q/N \sim 0.4$.

\par In one hand, the difference in $P_0^f$ between the Axelrod model and the analytical approach below $Q/N \sim 0.4$ can be explained through the fragmentation of the system during dynamics. 
Although the topology of the system is a complete graph, we define a fragment as a group of agents connected by non-zero homophily links. When a group of agents adopts  orthogonal cultural states respect to another group, they act as two independent fragments.
Once these clusters appear, the Axelrod model has no mechanisms to join them again. 
For instance, bottom panel of figure (\ref{fig:transicion}) shows the multiplicity of fragments for $N = 1024$. As can be observed, the system begins made up by a unique fragment but ended up fragmented, which is seen as an increment in the multiplicity. When the system is fragmented, the hypothesis of a mean-field approach fails. Panel (b) also suggests an explanation of why the analytical approach shows a transition in $P_0$:  the critical value ($Q_c/N \sim 0.4$) coincides with the value of $Q/N$ at which the system is already fragmented at the initial state.

\par On the other hand, since every link inside a fragment has similarity equal to $1$ at the final state (otherwise the system will keep evolving), a given value of $P_0^f$ in Axelrod simulations corresponds to the number of links between fragments (which ended up with similarity equal to $0$), normalized by the total number of links in the system. This means that difference between these simulations and theoretical values should correspond to the missing links due to fragmentation during the dynamics. In order to test this hypothesis, we  modify the value of $P_0^f$ in theoretical calculations by adding the contributions due to the inter-fragments missing links from the knowledge of the fragment distribution at the final state. When doing this, the modified value of $P_0^f$ (dashed yellow curve) matches exactly the simulations, as can be seen in upper panel of Figure \ref{fig:transicion}.
An interesting result is that by modifying $P_0$ with the missed links of the final fragments, it produces a good agreement with simulations for the whole dynamical trajectories. 
We can call $\hat P_i$ to the analytical values of $P_i$ corrected by the missing links. If $P_{inter}$ is the fraction of zero-similarity links due to the missing inter-fragment links, this reads as:
\begin{equation}
    \begin{split}
    \hat{P}_0&=P_{inter}+(1-P_{inter})P_0 \\
    \hat{P}_1&=(1-P_{inter})P_1 \\
    \hat{P}_2&=(1-P_{inter})P_2 \\
    \hat{P}_3&=(1-P_{inter})P_3.
    \end{split}
    \label{eq:reescaling}
\end{equation}

\par
\begin{figure}[ht]
    \centering
    \includegraphics[width = \columnwidth]{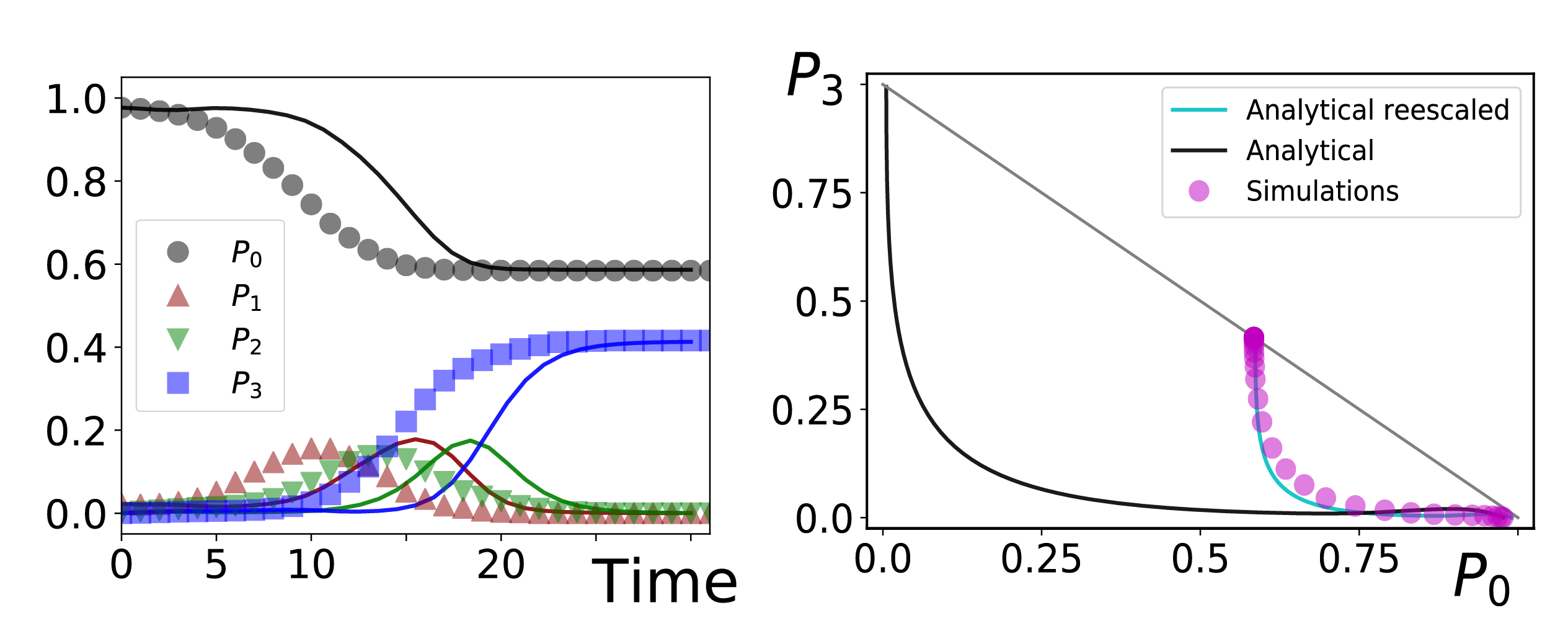}
    \caption{{\bf Axelrod and analytical transition in terms of $P_0^f$.} Re-scaling dynamic as time function (left) and trajectory (right) for $F=3$, $N=512$ and $Q=128$.}
    \label{fig:rescaling}
\end{figure}

 Figure \ref{fig:rescaling} shows that this modification produces an excellent approximation of the trajectory in the $P_0$-$P_3$ plane although is not so good when they are shown as function of time.
The idea behind the re-scaling proposed is that, if we would know the fragment distribution of the stationary state, we removed the inter-fragment links (that we know will end up with zero similarity) at $t=0$ as if they would not contribute during the whole dynamics.
Then, the analytical equations refers to the similarity distribution inside fragments, and the matching with simulations are recovered when combining with inter-fragment links of the stationary state through Eq.(\ref{eq:reescaling}).

\section{Discussion}

\par In this work, we present a novel mean-field approach of the Axelrod model based on similarity vectors on complete networks for $F=2$ and $F=3$.
In our analytical approach, once $F$ is set, the system depends on two parameters that play well-separated roles: On one hand, the parameter $Q$ (together with $F$) set the initial condition of the similarity distribution, while the system size $N$ couples the linear terms with cubic ones in master equations. 
\par Using this approach, we were able to exactly reproduce the dynamics of the similarity distribution for $F = 2$ for all value of $Q$, and correctly predict that, in this case, the dynamics does not depend on $N$ once the time step is set to $1/M$, where $M$ is the number of pairs of agents in the system. 
In the case of $F = 3$, our approach reproduces simulations while the system is made up by a unique fragment, condition that is satisfied in the thermodynamic limit for fixed $Q$ ($Q/N \to 0$).
During the Axelrod transition, the system can break up into groups of agents with orthogonal cultural states and the model has non mechanism to join them again. When this happens, an irreversible amount of links with similarity zero ($P_{inter}$) is created, the mean-field hypothesis does not hold, and the analytical approach fails. However, by a re-scaling of the similarity distribution which involves $P_{inter}$, the trajectories of simulations in phase diagrams can be recovered from the analytical equations. 
\par Specifically for the analytical system at $F = 3$, we found a  transition for large $N$ at $Q/N \sim 0.4$, where the stationary solution characterized by $P_0^f$ jumps from $0$ to $1$.
This values can be respectively identified with the monocultural and multicultural phases of the Axelrod model. We do not provide here a theoretical explanation of why this transition lies around $0.4$, but we think that this can be found by exploring the competition between the eigenvalues and eigenvectors of the linear system and the non-linear system (made up by only the cubic terms). 

\par Our work is not the first in proposing master-equations for the Axelrod model and follows past attempts like \cite{castellano2000nonequilibrium, vazquez2007non}. In all cases, the idea is to write a master-equation for the similarity distribution $P_m$, which in general looks like:
\begin{equation*}
    \frac{dP_m}{dt}=\sum_{r=1}^{F-1}\frac{r}{F}P_k \Big[ \delta_{m,r+1}-\delta_{m,r}+ (g-1)\sum_{n=0}^F(P_n W_{n,m}^{(r)}-P_m W_{m,n}^{(r)}) \Big]
\end{equation*}
where $g$ is the coordination number of the underlying network and $W_{n,m}^{(r)}$ are transition rates that take into account the probability of an indirect change for $n$ to $m$ due to a direct change with $r$ shared cultural features.
\par The main difference between analytical approaches lies in the calculation of $W_{n,m}^{(r)}$. Although useful insights about the Axelrod transition can be extracted for both \cite{castellano2000nonequilibrium} and \cite{vazquez2007non}, in these approaches correlations among adjacent links are neglected in order to make equations analytical tractable, leading to an inaccurate description of the dynamics of the system. In particular, in \cite{vazquez2007non} the transition rates involve a parameter $\lambda$ (interpreted as the conditional probability that two agents, $i$ and $k$, share a feature that is simultaneously not shared with a third one $j$), that is approximated by $\lambda = (Q-1)^{-1}$, involving the parameter $Q$ in the dynamics which we show can be exactly decoupled from it.
In our approach, correlations among agents are explicitly taken into account by writing the possible combination of similarity vectors.  
\par Another important difference in our approach is that when an agent $i$ copy a feature from $j$ changing its cultural state, we also update the $N-2$ similarities  defined between $i$ and any other agent $k \neq j$, no matter if the pair $i-k$ is topologically connected or not. Although in this work we consider a complete network where every pair of agents is connected, it is an important feature that must be considered if this approach is extrapolated to other network topologies. In \cite{castellano2000nonequilibrium, vazquez2007non}, only connected links are taken into account, which implies the presence of the coordination number $g$ in the general form of the master equation given above.
\par Finally, although we restrict our analysis to the cases $F=2$ and $F=3$, our approach can be seen as an algorithm to figure out all possible combinations of similarity vectors that can be translated into master-equations for larger values of $F$.

\appendix

\section{The Axelrod model and the similarity vector approach}

\par
\begin{figure}[ht]
    \centering
    \includegraphics[width = \columnwidth]{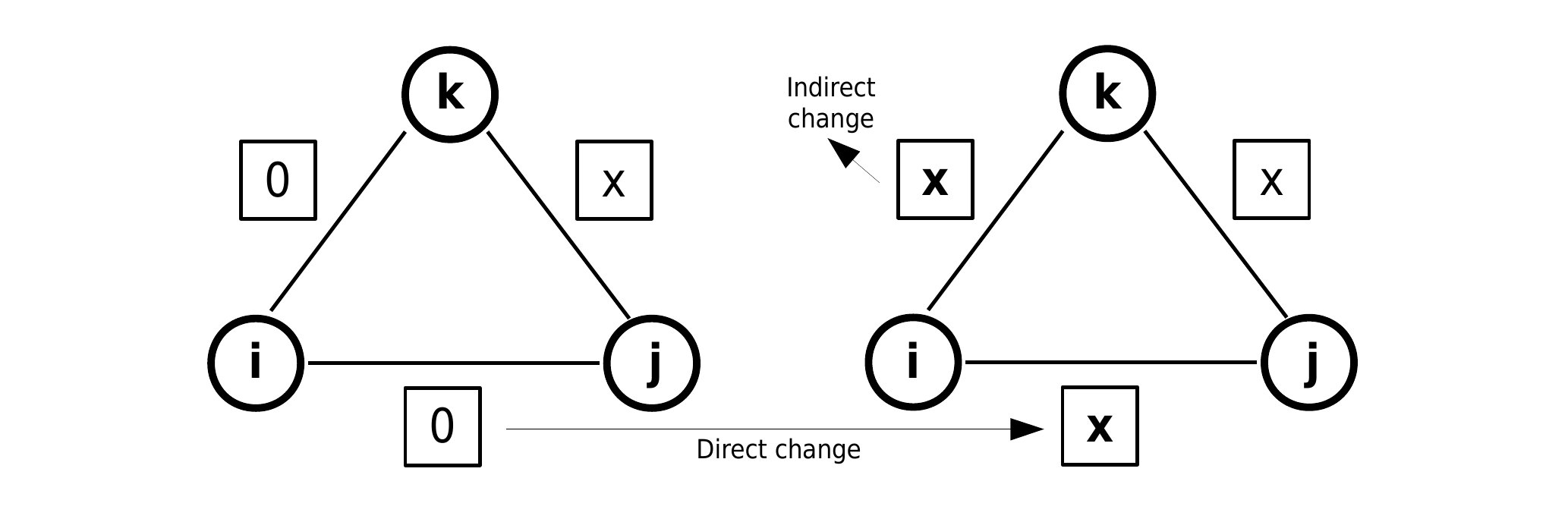}
    \caption{{\bf Dynamical behaviour in terms of similarity vectors.} 
    Suppose that one of the similarity features between agents $i$ and $j$ is changed from $0$ to $X$ by a direct change, and that it occurs when agent $i$ copies a cultural feature from $j$. Since only the cultural state of $i$ changes, the similarity feature between $j$ and $k$ remains constant. Then, the respective feature between $i$ and $k$ must necessarily change and adopt the same value that the similarity feature between $j$ and $k$ in order to avoid the banned state described in the main text.}
    \label{fig:Similarity_dynamic}
\end{figure}

 To reproduce the same dynamics than the Axelrod model, the dynamical rules must be rewritten in terms of the similarity vectors.
Figure \ref{fig:Similarity_dynamic} shows an example of the effects of changing the value of one similarity feature.
The most important fact is that, given three agents (which in this case means three similarity vectors), if we look at a specific similarity feature there is one banned state: To take the value of $X$ in two vectors and $0$ in the last one. 
For instance, consider figure \ref{fig:Similarity_dynamic} again: When the similarity feature of $i-j$ changes to $X$, if the feature $i-k$ doesn't change this would imply that agents $i$ and $j$ agree in that cultural feature, $j$ and $k$ do the same, but $i$ and $k$ not, which is a contradiction. 
\par Summarizing, the dynamical rules of the Axelrod model in terms of similarity vectors are the following:
\begin{itemize}
    \item Take a similarity vector of two connected agents, $i$ and $j$. With a probability proportional to the number of $X$ in their vector (which is the same as the homophily in the Axelrod model), change a random feature with value $0$ to $X$.
    \item Now, suppose that the change in the similarity vector comes from implicitly changing the value of a cultural feature in the state of $i$. Then, for all $k \neq i,j$, set the value of the similarity feature $i-k$ equal to the respective feature in $j-k$.
\end{itemize}
Finally, given the similarity vector $i-j$, if a given feature adopts the value of $X$, then the respective feature in the similarity vectors $i-k$ and $j-k$ must be equal for all $i, j, k$.

\par
\begin{figure}[ht]
    \centering
    \includegraphics[width = 0.9\textwidth]{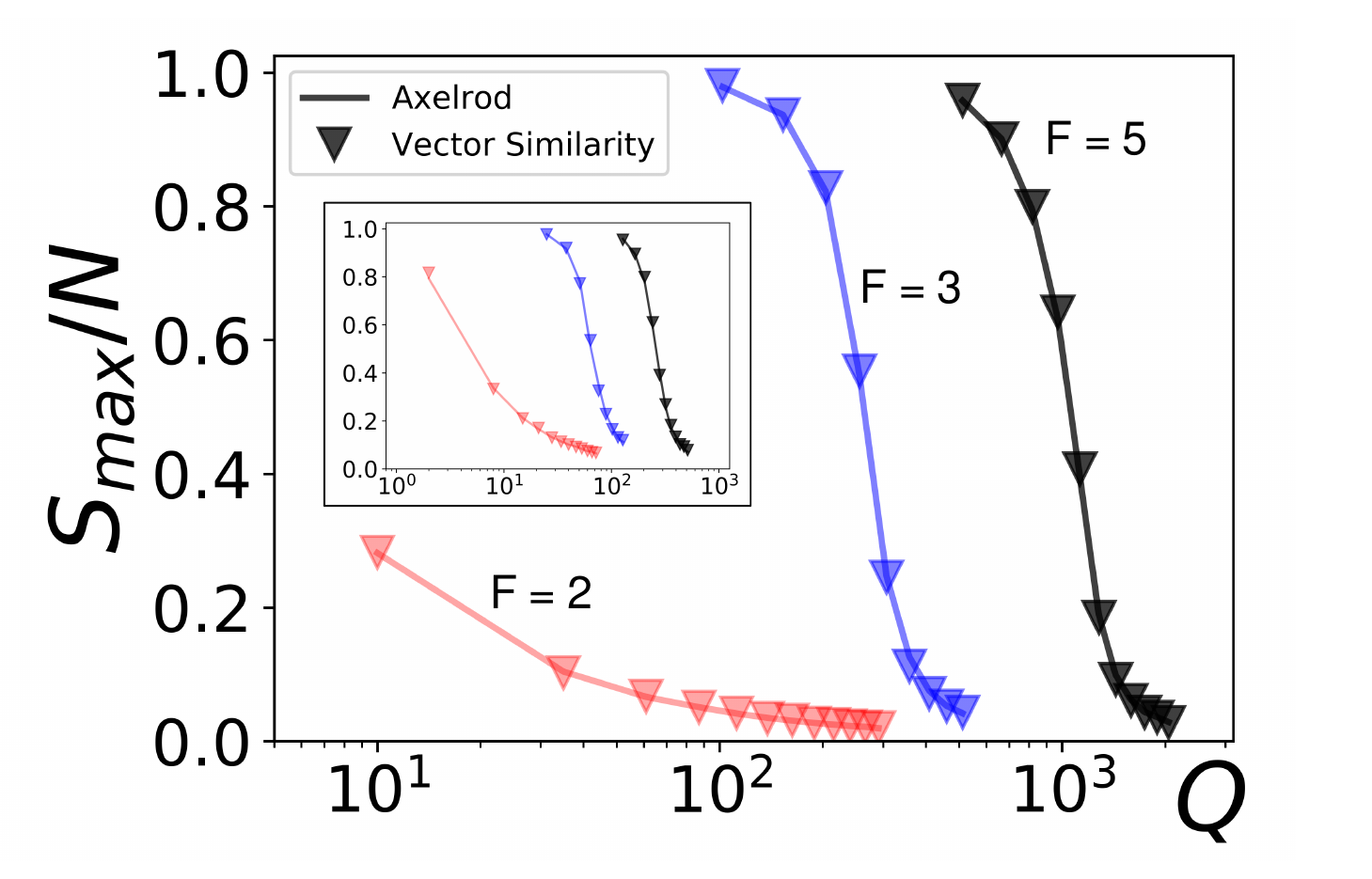}
    \caption{Axelrod model described by both cultural (full lines) and similarity vectors (triangles) on a complete network for $N = 1024$ agents and different values of $F$. Inset shows same results for $N = 256$. This figure shows that both representations are equivalent.}
    \label{fig:Axelrod_comparison}
\end{figure}

 Figure \ref{fig:Axelrod_comparison} shows the relative size of the biggest fragment at the final state $S_{max}/N$, in the Axelrod model in terms of both cultural and similarity vectors. $S_{max}/N$ is the usual observable to characterize the Axelrod transition. As figure shows, there are no significant differences in choosing one representation or the other.

\section{Derivation of $F = 3$ case master equations}
\label{sec:f3}

\par For $F=3$, let $P_0$ be the proportion of the total of the links that are the similarity vector $[0,0,0]$, $P_{1a}$, $P_{1b}$, $P_{1c}$ the three vectors for a link with one feature in common, $P_{2a}$, $P_{2b}$, $P_{2c}$ with two and $P_3$ the vector for a link that joins two equal states. Similarly to the $F=2$ we write all the feasibly terns where a direct change made an indirect one. Figure \ref{fig:EsquemaF3} show two examples for direct and indirect changes. The equation for the dynamic of these states is:

\begin{figure}[ht]
    \centering
    \includegraphics[width = \textwidth]{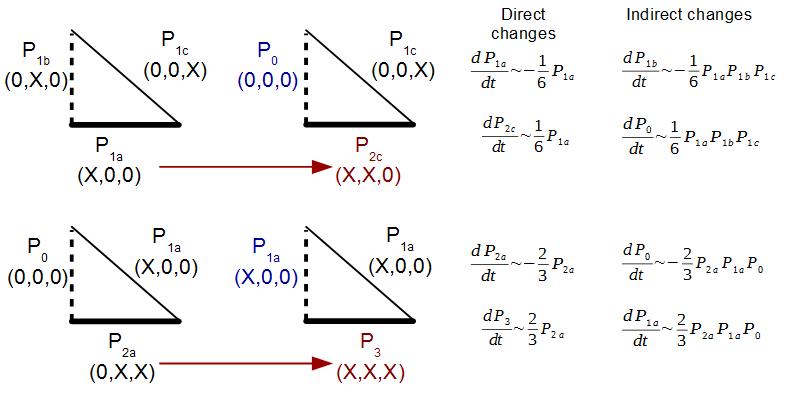}
    \caption{{\bf Example of dynamics of vector similarity states.}
    Two examples of links updates which involves direct and indirect interactions. Note that direct changes lead to linear terms in Eq.(\ref{eq:F3}), and these direct changes imply indirect ones, pointed out by dashed lines, which at the same time lead to cubic terms
    The $1/6$ and $2/3$ factor corresponds to the probability that a direct change effectively occurs (i.e., $1/F$ and $2/F$ for $F=3$ and one or two cultural feature shared) and the probability that the feature changed is the illustrated one (i.e., $1/2$ in the first example where there are two possible features to change in the direct interaction, and $1$ for the second example)}
    \label{fig:EsquemaF3}
\end{figure}

\begin{equation*}
\begin{split}
    \frac{dP_0}{dt} =&\frac{(N-2)}{3}\Big(-P_0(P_{1c}P_{2c}+P_{1b}P_{2b}+P_{1a}P_{2a})+3P_{1a}P_{1b}P_{1c}\Big) \\
    \frac{dP_{1a}}{dt} =&-\frac{P_{1a}}{3} + \frac{(N-2)}{3}\Big(-P_{1a}^2P_3 - P_{1a}P_{1b}P_{1c} + P_{1a}P_{2b}P_{2c} + \\ &+P_0P_{1b}P_{2b} + P_0P_{1c}P_{2c} - \frac{P_{1a}^2(P_{1c} + P_{1b})}{2}\Big)\\
    \frac{dP_{2a}}{dt} =& \frac{(P_{1b}+P_{1c})}{6}-2\frac{P_{2a}}{3} +
    \frac {(N-2)}{3}\Big(-P_{1a}P_{2a}P_0-P_{2a}(P_{1b}P_{2c}+P_{2b}P_{1c})+\\
    &+\frac{P_{1b}P_{1c}(P_{1b}+P_{1c})}{2}+P_3(P_{1b}^2+P_{1c}^2)\Big)\\
    \frac{dP_3}{dt}=& 2\frac{P_{2a}+P_{2b}+P_{2c}}{3} + \frac{(N-2)}{3}\Big(-P_3(P_{1a}^2+P_{1b}^2+P_{1c}^2)+P_{2a}P_{1b}P_{2c}+\\
    &+ P_{2a}P_{1c}P_{2b}+P_{2b}P_{1a}P_{2c}\Big)
\end{split}
\end{equation*}

\par Finally, we assume that different states with the same amount of features in common remain symmetric, that is $P_{1a}=P_{1b}=P_{1c}=\frac{P_1}{3}$ and $P_{2a}=P_{2b}=P_{2c}=\frac{P_2}{3}$, obtaining equations: 

\begin{equation*}
\begin{split}
    \frac{dP_0}{dt}&=\frac{(N-2)}{27}\Big(P_1^3-3 P_0 P_1 P_2\Big)\\
    \frac{dP_1}{dt}&=-\frac{P_1}{3}+\frac{(N-2)}{27}\Big(-2  P_1^3 -3 P_1^2 P_3 +P_1 P_2^2  +6 P_0 P_1 P_2  \Big)\\
    \frac{dP_2}{dt}&=\frac{P_1}{3}-\frac{2 P_2}{3}+\frac{(N-2)}{27}\Big(P_1^3+6 P_1^2 P_3-2 P_1 P_2^2 -3P_0 P_1 P_2\Big)\\
    \frac{dP_3}{dt}&=\frac{2 P_2}{3}+\frac{(N-2)}{27}\Big(-3 P_1^2 P_3 + P_1 P_2^2 \Big)
\end{split}
\end{equation*}

\section{Fixed points and stability analysis of $F = 3$ case master equations}

As mentioned in the main text, in addition to the set of fixed points when $P_1 = P_2 = 0$, this system has the following isolated fixed point:
\begin{equation*}
\begin{split}
P_0&=-2 \pm \frac{2}{3}\sqrt{9-\frac{2}{3c}}\\
P_1&=3 \mp \sqrt{9-\frac{2}{3c}}\\
P_2&=-\frac{3}{2} \pm \frac{1}{2}\sqrt{9-\frac{2}{3c}}\\
P_3&=\frac{3}{2} \pm \frac{1}{3}\sqrt{9-\frac{2}{3c}}
\end{split}
\end{equation*}
where $c = (N-2)/27$.
However, these point is an unfeasible one due to $P_1=-2P_2$ and therefore one of them is necessarily negative number.
\par By studying the linearized system around $P_1 = P_2 = 0$, we obtain the matrix:
\begin{equation*}
    L=
\begin{bmatrix} 
0 & 0 &0& 0\\
0 &-\frac{1}{3} & 0&0\\
0 &\frac{1}{3} & -\frac{2}{3}&0\\
0 &0 &\frac{2}{3}&0 
\end{bmatrix}
\end{equation*}
whose eigenvalues and eigenvectors are:
\begin{equation*}
\begin{split}
\lambda_1&=-2/3 \quad v_1=(0,0,-1,1)\\
\lambda_2&=-1/3 \quad v_2=(0,-1/2,-1/2,1)\\
\lambda_3&= 0  \quad  \quad \quad v_3=(0,0,0,1) \text{ y } v_4=(1,0,0,0)\\
\end{split}
\end{equation*}

\section{Stability of the non-linear part}

\par
\begin{figure}[ht]
    \centering
    \includegraphics[width = \columnwidth]{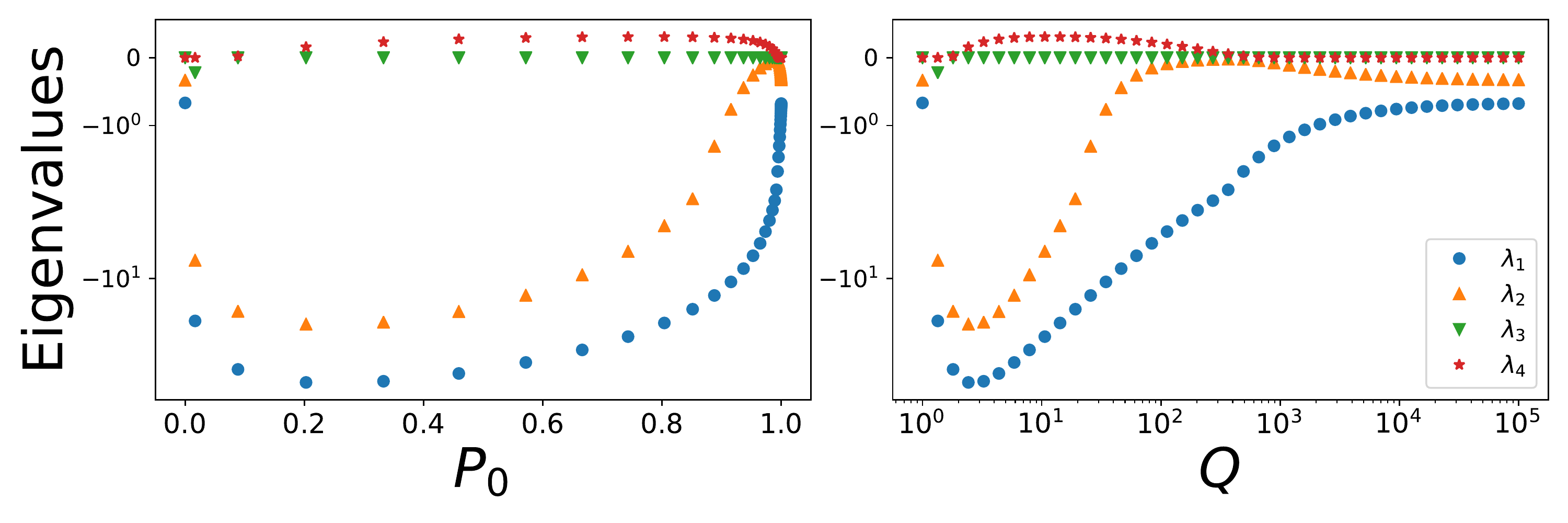}
    \caption{{\bf Eigenvalues for the linearized system of the non-linear terms.} Left figure as function of $P_0 = (1 - 1/Q)^3$, right figure as function of $Q$.}
    \label{fig:Eigenvalues}
\end{figure}

 By taking only the non-linear terms, we can calculate the first order matrix and evaluate it around initial conditions. The linearized matrix is read as:

\begin{equation*}
    D=
\begin{bmatrix} 
-3P_1P_2 & 3P_1^2-3P_0P_2 &-3P_0P_1& 0\\
6P_2P_1 &-6P_1^2-6P_1P_3+6P_0P_2+P_2^2 & 6P_0P_1+2P_2P_1 & -3P_1^2\\
-3P_1P_2 &-3P_0P_2-2P_2^2+12P_3P_1+3P_1^2 & -3P_0P_1-4P_2P_1&6P_1^2\\
0 &-6P_3P_1+P_2^2 &2P_2P_1&-3P_1^2 
\end{bmatrix}
\end{equation*}
We numerically calculate the eigenvalues of this matrix at the initial condition. Figure \ref{fig:Eigenvalues} shows these values as function of the initial $P_0$. As we can see in this figure, the most dominant eigenvalue has negative sign, leading that the initial conditions are also a stable manifold.

\section{Null non-linear terms at initial condition}

\par We can rewrite Eq.(\ref{eq:F3}) by defining $A=P_1^2-3 P_0 P_2$ and $B=P_2^2-3 P_1 P_3$:
\begin{equation*}
\begin{split}
    \frac{dP_0}{dt}&=\frac{(n-2)}{27} P_1 A\\
    \frac{dP_1}{dt}&=-\frac{P_1}{3}+\frac{(n-2)}{27}P_1(B-2A)\\
    \frac{dP_2}{dt}&=\frac{P_1}{3}-\frac{2 P_2}{3}+\frac{(n-2)}{27}P_1(A-2B)\\
    \frac{dP_3}{dt}&=\frac{2 P_2}{3}+\frac{(n-2)}{27}P_1 B
\end{split}
\end{equation*}

At a given $Q$, the initial similarity distribution is:
\begin{equation*}
\begin{split}
P_0&=\Big(1-\frac{1}{q}\Big)^3\\
P_1&=3\Big(1-\frac{1}{q}\Big)^2\frac{1}{q}\\
P_2&=3\Big(1-\frac{1}{q}\Big)\Big(\frac{1}{q}\Big)^2\\
P_3&=\Big(\frac{1}{q}\Big)^3
\end{split}
\end{equation*}
In this case both $A$ and $B$ are null:
$$A=9\Big(1-\frac{1}{q}\Big)^4\Big(\frac{1}{q}\Big)^2-3 \Big(1-\frac{1}{q}\Big)^3 \Big[3\Big(1-\frac{1}{q}\Big)\Big(\frac{1}{q}\Big)^2\Big]=9\Big(1-\frac{1}{q}\Big)^4\Big(\frac{1}{q}\Big)^2-9\Big(1-\frac{1}{q}\Big)^4\Big(\frac{1}{q}\Big)^2=0$$

$$B=9\Big(1-\frac{1}{q}\Big)^2\Big(\frac{1}{q}\Big)^4 - 3 \Big[3\Big(1-\frac{1}{q}\Big)^2\frac{1}{q}\Big] \Big(\frac{1}{q}\Big)^3=0$$
Therefore, at the initial condition, non-linear terms are null and the dynamics in only driven by the linear ones.


\begin{thebibliography}{10}

\bibitem{axelrod1997dissemination}
Robert Axelrod.
\newblock The dissemination of culture: A model with local convergence and
  global polarization.
\newblock {\em Journal of conflict resolution}, 41(2):203--226, 1997.

\bibitem{battiston2017layered}
Federico Battiston, Vincenzo Nicosia, Vito Latora, and Maxi San~Miguel.
\newblock Layered social influence promotes multiculturality in the axelrod
  model.
\newblock {\em Scientific reports}, 7(1):1--9, 2017.

\bibitem{castellano2000nonequilibrium}
Claudio Castellano, Matteo Marsili, and Alessandro Vespignani.
\newblock Nonequilibrium phase transition in a model for social influence.
\newblock {\em Physical Review Letters}, 85(16):3536, 2000.

\bibitem{hernandez2018robustness}
Alexis~R Hern{\'a}ndez, Carlos Gracia-L{\'a}zaro, Edgardo Brigatti, and Yamir
  Moreno.
\newblock Robustness of cultural communities in an open-ended axelrod’s
  model.
\newblock {\em Physica A: Statistical Mechanics and its Applications},
  509:492--500, 2018.

\bibitem{klemm2003nonequilibrium}
Konstantin Klemm, V\'ictor~M Egu\'iluz, Ra\'ul Toral, and Maxi San~Miguel.
\newblock Nonequilibrium transitions in complex networks: A model of social
  interaction.
\newblock {\em Physical Review E}, 67(2):026120, 2003.

\bibitem{klemm2005globalization}
Konstantin Klemm, V\'ictor~M Egu\'iluz, Raul Toral, and Maxi San~Miguel.
\newblock Globalization, polarization and cultural drift.
\newblock {\em Journal of Economic Dynamics and Control}, 29(1-2):321--334,
  2005.

\bibitem{lanchier2016fixation}
Nicolas Lanchier and Paul-Henri Moisson.
\newblock Fixation results for the two-feature axelrod model with a variable
  number of opinions.
\newblock {\em Journal of Theoretical Probability}, 29(4):1554--1580, 2016.

\bibitem{lanchier2013fixation}
Nicolas Lanchier, Stylianos Scarlatos, et~al.
\newblock Fixation in the one-dimensional axelrod model.
\newblock {\em The Annals of Applied Probability}, 23(6):2538--2559, 2013.

\bibitem{maccarron2020agreement}
P{\'a}draig MacCarron, Paul~J Maher, Susan Fennell, Kevin Burke, James~P
  Gleeson, Kevin Durrheim, and Michael Quayle.
\newblock Agreement threshold on axelrod’s model of cultural dissemination.
\newblock {\em Plos one}, 15(6):e0233995, 2020.

\bibitem{morris2019impact}
Rhodri Morris, Liam Turner, Roger Whitaker, and Cheryl Giammanco.
\newblock The impact of peer pressure: Extending axelrod's model on cultural
  polarisation.
\newblock In {\em 2019 IEEE International Conference on Cognitive Computing
  (ICCC)}, pages 114--121. IEEE, 2019.

\bibitem{pinto2020erdos}
Sebasti{\'a}n Pinto and Pablo Balenzuela.
\newblock Erd{\'o}s-r{\'e}nyi phase transition in the axelrod model on complete
  graphs.
\newblock {\em Physical Review E}, 101(5):052319, 2020.

\bibitem{stivala2016another}
Alex Stivala and Paul Keeler.
\newblock Another phase transition in the axelrod model.
\newblock {\em arXiv preprint arXiv:1612.02537}, 2016.

\bibitem{vazquez2007non}
Federico V{\'a}zquez and Sidney Redner.
\newblock Non-monotonicity and divergent time scale in axelrod model dynamics.
\newblock {\em EPL (Europhysics Letters)}, 78(1):18002, 2007.

\bibitem{vilone2002ordering}
Daniele Vilone, Alessandro Vespignani, and Claudio Castellano.
\newblock Ordering phase transition in the one-dimensional axelrod model.
\newblock {\em The European Physical Journal B-Condensed Matter and Complex
  Systems}, 30(3):399--406, 2002.

\end{thebibliography}
\end{document}